\documentclass[
aip,
rsi,
twocolumn,
10pt
]{revtex4-1}

\usepackage[hidelinks,
colorlinks=true,
linkcolor=blue,
citecolor=blue,
urlcolor=blue]{hyperref}
\usepackage{graphicx}
\usepackage[english]{babel}
\usepackage{color}

\begin{document}

\title{\color{blue} \Large Note: Low-noise high-voltage DC power supply for nanopositioning applications}

\author{Cristian H. Belussi}
\email{cristian.belussi@cab.cnea.gov.ar}
\affiliation{Low Temperature Division and Instituto Balseiro, Centro At\'omico
Bariloche, 8400 Bariloche, Argentina.}

\author{ Mariano G\'omez Berisso}
\affiliation{Low Temperature Division and Instituto Balseiro, Centro At\'omico
Bariloche, 8400 Bariloche, Argentina.}

\author{Yanina Fasano}
\affiliation{Low Temperature Division and Instituto Balseiro, Centro At\'omico
Bariloche, 8400 Bariloche, Argentina.}

\date{\today}

\begin{abstract}
Nanopositioning techniques currently applied  to characterize
physical properties of materials interesting for applications at
the microscopic scale rely on high-voltage electronic control
circuits that should have the lowest possible noise level. Here we
introduce a simple, flexible, and custom-built power supply
circuit that can provide +375\,V with a noise level below 10\,ppm.
The flexibility of the circuit comes from its topology based on
discrete MOSFET components  that can be suitable replaced in order
to change the polarity as well as the output voltage and current.
\end{abstract}



\maketitle

The study of electronic and magnetic properties of novel materials at the atomic
scale rely on developing nanopositioning techniques with low-noise
level. In several fields, such as condensed matter and optical
devices, the nanopositioning of the probes is implemented by
means of highly-capacitive piezoelectric motors (tens of nF) \cite{Pohl1987}. In
order to improve the stability of the positioning system the
high-voltage power supply (200-600\,V) feeding these motors should
have the lowest possible noise level. The works available in the
literature \cite{ChenX1992,Usher1993,Colclough2000,Muller2005,Flaxer2006,ChenL2012a}
on the control electronics of piezoelectric motors present low
level of detail regarding the design, implementation, and
characterization of such high-voltage power supplies.

In this note we provide a comprehensive report on a low-noise
 level power-supply for the control electronics of piezoelectric motors that
has the advantages of simplicity and flexibility. We present and
characterize a custom-made +375\,V, high-current (0.1\,A) and
low-noise ($< 10$\,ppm) performance power supply successfully
integrated onto the control electronics \cite{Belussi2013} of a
piezoelectric motor with sub-Angstrom positioning resolution.

\begin{table}[bbb]
  \centering
  \begin{tabular}{|c|c|}
  \hline
  Output voltage & +\,375V $\pm$\, 1\,\% \\ \hline
  Output current & 30\,mA \\ \hline
  Noise ripple & $<$ 7\,mV$_{rms}$  \\ \hline
  Efficiency & $>$ 85\% \\ \hline
  Load regulation & $<$ 0.03\% \\ \hline
  Short-circuit time &  $\infty$ \\
  \hline
\end{tabular}
  \caption{Specifications of the +\,375\,V low-noise power supply.}
  \label{tampgps}
\end{table}

\begin{figure*}[ttt,flaotfix]
\includegraphics[width=0.85\textwidth]{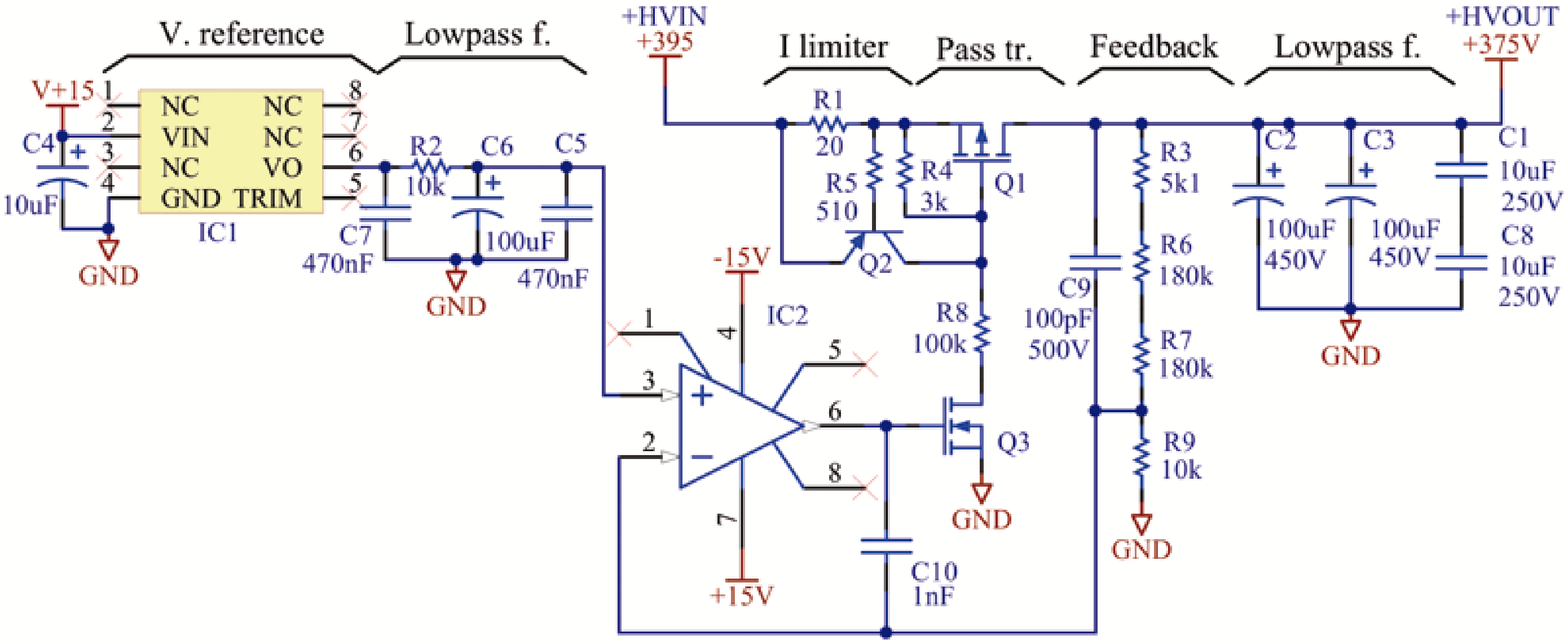}
\caption{Detailed circuit for the +\,375\,V  high-voltage
low-noise power supply. The transistors are Q1:MTP2P50; Q2:BC557C;
Q3:BUZ80A. Resistances are given in Ohm and have a power
dissipation rating of 0.5\,W. Capacitances have a voltage rating
of 25\,V, except indicated otherwise.} \label{FuenteG}
\end{figure*}

Well known power-supply design techniques allow the implementation
of high output voltages using for instance integrated operational
amplifiers \cite{ApexAnalog}, high-voltage regulators
\cite{Flaxer2006}, low-voltage regulators \cite{Flaxer2008},
switching converters\cite{Ertl1997a}, voltage multipliers
\cite{ChenL2012a} and discrete amplifiers\cite{Kocum2011}.
However, applying these solutions in order to satisfy our
technical requirements is not straightforward. Switching
converters can provide up to 1\,kV but present some drawbacks as
high level of electromagnetic interference, high-frequency noise,
load-dependent ripple, and the isolation problems associated with
operating a high-frequency transformer \cite{Williams2008}. In the
case of voltage multipliers based on arrays of capacitors and
diodes, loading with a high-current demanding circuit results in a
considerable level of ripple on the DC output voltage
\cite{Beck2008}. Zener diodes used as voltage references without a
proper compensation, as well as discrete circuits without feedback
loop, present drift problems. Discrete amplifiers like
common-source and common-drain stages use  MOS-N transistors
requiring a non-desirable high-power dissipation in order to be
polarized.

In our case we solve this problem by means of the linear
low-dropout regulator circuit \cite{LDO_TI_1999, ldo_ti_stab} shown in Fig.\,\ref{FuenteG}
providing a DC voltage output of +375\,V. This circuit has a low-noise voltage
 reference (IC1), a low-pass filter (C5, C6, C7, R2), a low-noise operational amplifier
(IC2), and high-voltage transistors (Q1, Q3). The load regulation
and ripple of  the rectified wave are controlled by the feedback
loop between the operational amplifier and the resistors network.
The high-voltage source that feeds the circuit comes from a
custom-built step-up transformer and a full-wave rectifier bridge
with filter capacitors (not shown in the figure) providing +395\,V
to the regulator. The circuit regulates the output voltage to a
fixed value generated by the reference voltage times the feedback
gain, with the pass transistor Q1 operating in the active region
of the MOSFET \cite{LDO_TI_1999}. Adequate values for  R4 and R8
must be selected accordingly to the V$_{\rm{GS}}^{\rm{ON}}$ values
of the Q1 transistor and to the leakage currents in the
polarization network\cite{LDO_TI_1999}. The efficiency of the
circuit is of 85\% in normal operation at 30\,mA . The main
specifications of our power-supply are summarized in table
\ref{tampgps}.

In order to reduce the power-supply output noise we have chosen
the low-noise voltage reference LT1021 providing +\,10\,V in the
regulator with a noise level lower than 1\,ppm in 10\,Hz and a
thermal-drift smaller than 5\,ppm/K. The low-pass network was
implemented with a cut-off frequency of 1\,Hz in order to minimize
noise from the reference \cite{Maxim2005,Nogawa2012}. With the aim of reducing the noise
introduced in this stage we used a NE5534 operational amplifier.
An additional advantage of the low-dropout topology of our circuit
is the low power required to polarize the transistors, and the possibility
of high output currents with high efficiency \cite{LDO_TI_1999}.

\begin{figure}[ttt, flaotfix]
  \centering
  \includegraphics[width=0.35\textwidth]{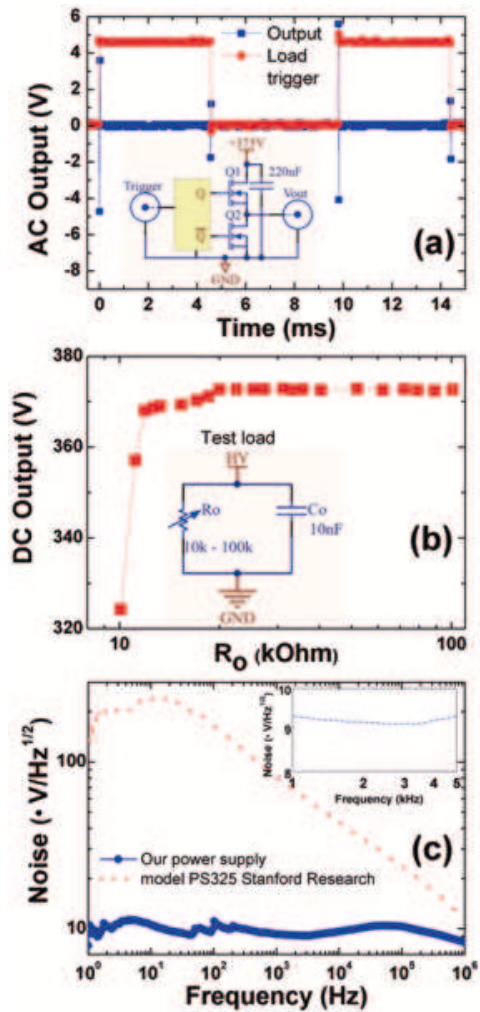}
  \caption{Figures of merit of the +375\,V low-noise power supply.
  (a) AC output voltage (blue squares) of the power supply
  when connected and disconnected of a load of 10\,nF in parallel with
  1\,M$\Omega$. The switching on and off of the load is perform
  via triggering (red dots) the bridge circuit shown in the insert. (b) DC output voltage  regulation with
  a variable load of 10 to 100\,k$\Omega$ in parallel with 10\,nF. (c) Noise spectrum up to 1\,MHz
  with an output load of 1\,M$\Omega$ in parallel with 10\,nF for our power supply (full points) and the commercial
  PS325 model from Stanford Research (dotted line). In both cases, data were acquired by means of a DSP lock-in amplifier
  with a time constant of 50\,ms. Insert: detail of the noise level in our circuit in the typical frequency range for
  driving piezoelectric motors.}
  \label{GVOZ}
\end{figure}

The circuit presented here is rather flexible and can be easily
adapted to implement power supplies with lower and higher output
voltages as well as inverse polarity. A negative voltage-output
supply can be implemented with the same elements working in
inverse polarity and replacing the MOS-N  with a MOS-P transistor,
and viceversa \cite{LDO_TI_1999}.
The absolute value of the voltage output can be
increased up to $\pm$\,600\,V by conveniently choosing the feedback
resistors. The practical limit for the largest absolute voltage
output of the source  is the maximum V$_{\rm{DS}}$ across the
    transistors.  For the circuit proposed here the transistors are
    MOS-P MTP2P50 (Q1) and MOS-N BUZ80A (Q3), withstanding up to 500
and 800\,V, respectively.  Alternatively, the MOS-P transistor can
be replaced by the IXTH16P60P component withstanding a maximum of
600\,V and a continuous current of 16\,A. Concerning the
efficiency of the circuit, if  higher power or power-factor are
required, a switching converter can be added as a previous stage
to step-up the voltage, in place of the rectifier and transformer.

The design of this circuit considered faults due to
short-circuit or current overload with the current limit set by
R1.  In order to tolerate an infinite-time short-circuit the
heat-sink of the pass-transistor (Q1) has to be chosen carefully
considering that in our design the maximum current is of 30\,mA
and that the transistor works for voltages smaller than +\,375\,V.
This current limit can be increased up to the maximum drain-source
current that for Q1 is of  2\,A. For this current-limiting stage,
we connected the Q2 transistor in a floating  configuration
avoiding the use of a high-voltage component.

One application of this power supply is the feeding of the control
electronics of piezoelectric motors that for coarse motion work
normally in a slip-stick mode \cite{Chatterjee2009, Renner1990}.
Therefore we tested its stability during the transient regime of a
capacitive load of the order of piezoelectric motors excited with
asymmetric signals. This was performed by switching on and off
 the bridge circuit shown in the insert to
Fig.\,\ref{GVOZ}(a), connected between the source output and a
load of 10\,nF. The main panel of Fig.\,\ref{GVOZ}(a) shows the
transient and stationary response at 100\,Hz. On switching on and
off the circuit a voltage spike of only $\sim 2\,$\% of the
nominal output develops. Similar results are obtained up to
10\,kHz. This response is due to the abrupt discharge of the
capacitors of the source. This figure of merit is reasonably good
in order to abruptly charge and discharge a piezoelectric motor in
slip-stick coarse positioning mode.

In addition, we performed a load regulation test by using the load
shown in the insert to Fig.\, \ref{GVOZ} (b). The evolution of the
output voltage on varying the resistor $R_{0}$ is shown in the
main panel of the same figure. In the normal operation regime, the
output voltage is within 0.03\,\% of the nominal value. The mean
voltage in normal operation is 372.6\,V yielding an
 accuracy of 0,64\,\%. For impedances $R_{0} <
11$\,k$\Omega$ the maximum output current is reached and the
output voltage drops drastically.

The noise in the output voltage has been characterized with a
digital-signal-processing lock-in amplifier. The noise spectrum up
to 1\,MHz is presented in Fig.\,\ref{GVOZ} (c) in the case of a
10\,nF load in parallel with a 1\,M$\Omega$ resistor. The output
noise level in this large frequency range is below
12\,$\mu$V/$\sqrt{\rm{Hz}}$. The noise spectrum integrated in the
100\,kHz bandwidth yields an output noise lower than
2\,mV$_{\rm{peak}}$. This noise level is one order of magnitude
better than relatively expensive commercially-available external
high-voltage power supplies such as model PS325 of Stanford Research
\cite{StanfordR} (see dotted line in Fig.\,\ref{GVOZ}). Our power
supply was successfully integrated onto the control electronics
\cite{Belussi2013} that drives a scanning probe piezoelectric
motor with sub-Angstrom resolution.

The figures of merit and specifications of the circuit presented
here makes it a suitable choice to feed actuators, sensors, or to
power other high-voltage electronics in order to perform on
low-noise level laboratory experiments. Our circuit is based in
MOSFET technology components available in any local market that
can be easily and suitably replaced in order to provide different
polarities as well as output voltage and current values. In
conclusion, the circuit presented here is a simple,
easy-accessible and flexible custom-built solution for
high-voltage low noise level power supplies to be integrated onto
control electronics for nanopositioning applications.

\bibliography{bib_fuente}

\end{document}